\documentclass[hidelinks,conference,a4paper]{IEEEtran}
\usepackage{textcomp}
\usepackage{xcolor}
\usepackage{footnotebackref}
\usepackage{footmisc} 
\usepackage{mathtools}
\usepackage{multicol}
\usepackage{cuted}
\usepackage{lipsum, color}
\usepackage{tabularx,booktabs}
\usepackage{hhline}

\usepackage{colortbl}
\usepackage{color} 

\usepackage{balance}    
\usepackage{bigstrut}
\usepackage{cite}       
\usepackage{threeparttable}     
\newcolumntype{M}[1]{>{\centering\arraybackslash}m{#1}}  
\usepackage{color} 
\newcommand{\Abox}[2][6]{\makebox[#1pt][r]{#2}}  
\newcommand{\Aignore}[1]{}     
\newcommand{\Agraycell}{\cellcolor[rgb]{ .93,  .93,  .93}}

\usepackage{pgffor}

\IEEEoverridecommandlockouts        

\begin{document}

\title{FPGA Processor In Memory Architectures (PIMs):\\Overlay or Overhaul ?}


{
\author{
\IEEEauthorblockN{
    MD Arafat Kabir\IEEEauthorrefmark{1}, 
    Ehsan Kabir\IEEEauthorrefmark{1}, 
    Joshua Hollis\IEEEauthorrefmark{1}, 
    Eli Levy-Mackay\IEEEauthorrefmark{1},
    Atiyehsadat Panahi\IEEEauthorrefmark{2}, \\
    Jason Bakos\IEEEauthorrefmark{3},
    Miaoqing Huang\IEEEauthorrefmark{1} and
    David Andrews\IEEEauthorrefmark{1}
}
\IEEEauthorblockA{
Department of Computer Science and Computer Engineering \\
\IEEEauthorrefmark{1}%
    University of Arkansas, \ \ 
\IEEEauthorrefmark{3}%
    University of South Carolina, \ \ 
\IEEEauthorrefmark{2}%
    Cadence Design Systems
}

\{makabir, ekabir, jrhollis, elevymac, apanahi, mqhuang, dandrews\}@uark.edu, 
jbakos@cse.sc.edu%
\thanks{This work is partially supported by the National Science Foundation under Grant No. 1955820.}
}   
}   

\onecolumn
{\large\vspace*{\fill}

© 2023 IEEE.  Personal use of this material is permitted.  
Permission from IEEE must be obtained for all other uses, 
in any current or future media, including reprinting/republishing this 
material for advertising or promotional purposes, 
creating new collective works, for resale or redistribution to servers or lists, 
or reuse of any copyrighted component of this work in other works. \\

This work has been accepted at the 
2023 33rd International Conference on Field-Programmable Logic and Applications (FPL)
and will appear in the proceedings and on the IEEE website soon.

\vspace*{\fill}
}
\twocolumn

\maketitle

\begin{abstract}
The dominance of machine learning and the ending of Moore's law have renewed interests in Processor in Memory (PIM) architectures.  
This interest has produced several recent proposals to modify an FPGA's BRAM architecture to form 
a next-generation PIM reconfigurable fabric\cite{comefa,eriko}.  
PIM architectures can also be realized within today's FPGAs as overlays without the need to modify the underlying FPGA architecture.  
To date, there has been no study to understand the comparative advantages of the two approaches.  
In this paper, we present a study that explores the comparative advantages between two proposed 
custom architectures and a PIM overlay running on a commodity FPGA.   
We created PiCaSO, a \underline{P}rocessor \underline{i}n/near Memory S\underline{ca}lable and Fa\underline{s}t 
\underline{O}verlay architecture as a representative PIM overlay. 
The results of this study show that the PiCaSO overlay achieves up to 80\% of the peak throughput of the 
custom designs with 2.56$\times$ shorter latency and \mbox{25\% -- 43\%} better BRAM memory utilization efficiency.  
We then show how several key features of the PiCaSO overlay can be integrated into the custom PIM designs 
to further improve their throughput by 18\%, latency by 19.5\%, and memory efficiency by 6.2\%.

\end{abstract}


\section{Introduction}
Convolutional Neural Networks (CNNs), Multilayer Perceptrons (MLPs), and Recurrent Neural Networks (RNNs) have emerged as the dominant machine learning approaches for today's application domains.  Each of the three networks have different computation to communication requirements, or operational intensities, that necessitate different types of architectural support~\cite{tpu2017}. 

CNNs exhibit high operational intensities where end-to-end inference latencies are dominated by arithmetic compute times.  
Conversely, MLPs and RNNs exhibit much lower operational intensities where the end-to-end inference 
latencies are dominated by bus bandwidth and memory swapping times. 
Processor in/near memory (PIM) architectures~\cite{Kogge1997,terasys,stone,CompRAM} are making a resurgence to address these types of network requirements.  
PIM architectures break the sequential von Neumann bottleneck by integrating bit-serial processors within memory.  

PIM architectures can leverage the continued trend in machine learning arithmetic towards lower precision. 
Less than full precision operands can result in better utilization of limited memory, 
and the bit-serial processing elements (PEs) can provide better energy efficiency compared to full precision PEs.  
PIM systems offer a theoretical peak performance limited only by the memory bandwidth.

The trend towards PIM architectures is inspiring new reconfigurable fabrics that integrate bit-serial arithmetic units into BRAM IP to form PIM tiles~\cite{eriko,comefa,MicronIMI,PIMproto,samsung,eckert,ahn2015pim,deming,floatPIM2019,compDRAM2019}.  
These architectures may represent the future but are not currently available. To fill the void PIM architectures can be created as overlays in existing FPGAs.  
The fundamental question we explore in our work is,
how close in performance can an overlay come to the performance being reported for next-generation PIM reconfigurable compute fabrics? 

To explore this question we created PiCaSO, a \underline{P}rocessor \underline{i}n/near Memory S\underline{ca}lable and Fa\underline{s}t 
\underline{O}verlay as an open-source PIM overlay architecture~\cite{github_picaso}. 
We present performance comparisons that show PiCaSO achieves 80\% of the peak throughput of these emerging 
proposed custom designs while delivering 2.56$\times$ shorter latency and \mbox{25\% -- 43\%} better BRAM memory utilization efficiency. 
This validates PiCaSO's ability to bring enhanced designer productivity to the design 
of FPGAs without the traditional performance sacrifices of an overlay.

Finally, we apply several PiCaSO design optimizations to the custom PIM designs to further improve their 
throughput by 18\%, latency by 19.5\%, and memory efficiency by 6.2\%.
The specific contributions of this work are:

\begin{itemize}
\item 
A PIM overlay architecture that scales linearly with the BRAM capacity of a device, without sacrificing the clock frequency.

\item 
A comparative study with a state-of-the-art PIM overlay showing improvements of clock speed by 2$\times$, 
resource utilization by 2$\times$, and accumulation latency by 17$\times$.

\item
An improved version of an existing custom PIM design incorporating the novel features of the proposed overlay architecture.  

\item
A comparative study between the proposed overlay and custom PIM designs analyzing the trade-offs 
and use cases of the overlay and custom designs.

\end{itemize}
PiCaSO is open-source and freely available at~\cite{github_picaso} for use, modification, and distribution without restriction.

\section {Related Work}
\label{sec:back}
PIM architectures are a growing area of research\cite{MicronIMI,PIMproto,samsung,eckert,ahn2015pim,deming,floatPIM2019,compDRAM2019}.  
Building on earlier work such as Logic-In-Memory\cite{stone}, Terasys\cite{terasys}, Shamrock\cite{Kogge1997}, and Computational RAM\cite{CompRAM}, PIM architectures 
seek to break the classic von Neumann bottleneck by moving the processing closer to the data residing in memory in a  
Single Instruction Multiple Data (SIMD) architectural organization~\cite{stitt,philip1,csordas,landy,walsh,finn,stripes,lowprec}.

\begin{figure}
\centering
\includegraphics[width=\linewidth]{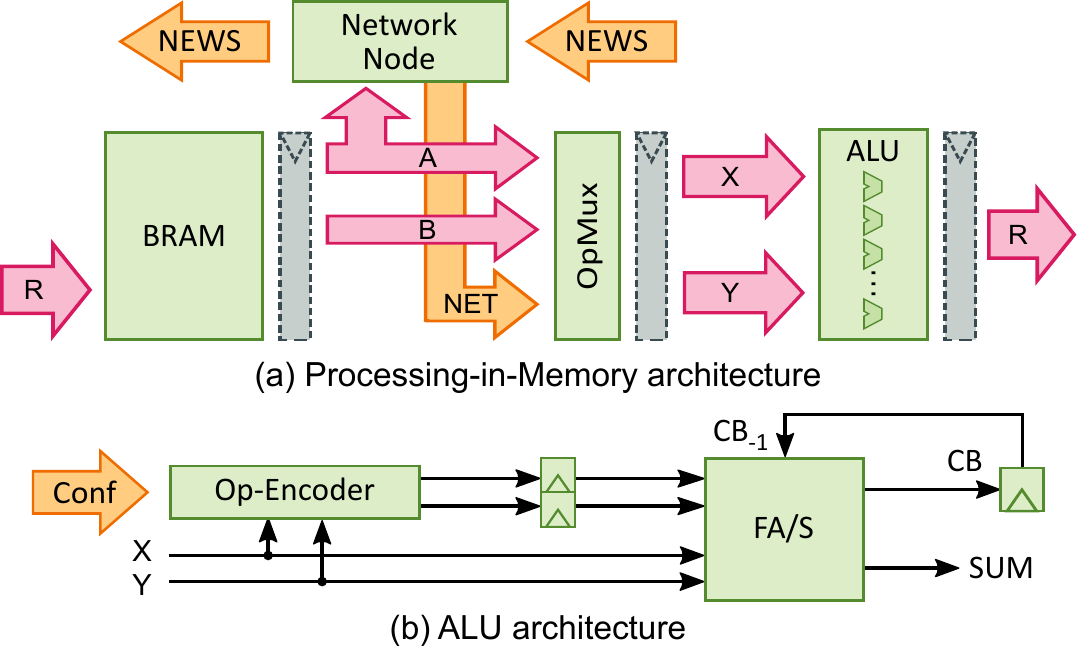}
\caption{Proposed overlay architecture for processing in/near memory, PiCaSO}
\label{fig:archPIM}
\end{figure}

Recently, the reconfigurable computing community has been exploring modifying the internal circuity of 
the on-chip BRAMs within a modern FPGA with bit-serial arithmetic and logic operations to form a PIM tile~\cite{comefa, eriko}. 
Examples include RIMA (Reconfigurable In-Memory Accelerator)~\cite{eriko}, which is built upon Neural Cache~\cite{eckert}.
Their compute capable BRAMs (CCB) enhanced the peak MAC throughput by factors of 1.6$\times$ and 2.3$\times$ for 8-bit integer 
and block floating point precision at a cost of a 7.4\% increase in BRAM tile area\cite{eriko}.  
Reported clock frequencies range from 250 MHz to 455 MHz on a Straix 10 device.  
CCB requires simultaneous activation of multiple wordlines on a port and modifications to the voltage source for robustness.

CoMeFa~\cite{comefa} builds upon CCB taking advantage of the dual-port nature of BRAMs.
Two versions of PIM blocks were proposed. 
Optimized for the delay, CoMeFa-D showed 25.4\% tile area increase due to 
the inclusion of 160 PEs, 120 sense amplifiers (SA), and write drivers.  
Optimized for the area, CoMeFa-A showed an 8.1\% increase in the BRAM tile area mainly attributed to the addition of 40 PEs.  
The maximum clock frequency (735 MHz) dropped 1.25$\times$ to 588 MHz for CoMeFa-D.  
The clock frequency for CoMeFa-A dropped 2.5$\times$ to 294 MHz, to perform 4 reads and 2 writes in a single cycle.

PiCaSO is very synergistic with these efforts.  
We show how design optimizations developed for PiCaSO can be applied to these BRAM tile designs
and potentially reclaim the clock frequency difference with the BRAM's supported maximum.

\section{PiCaSO Architecture}
\label{sec:proposed}

Fig.~\ref{fig:archPIM} shows the processor in-memory architecture of PiCaSO.  PiCaSO builds on the SPAR-2 PIM processor array reported in\cite{spar21,spar22,panahi22} but with the key modifications discussed below.
Custom bit-serial PIM designs including those reported in~\cite{spar22, comefa, eriko} stream operands 
between memory and ALUs across dedicated bitlines. 
Such an organization does not provide support for fast reduction operations (summation of product terms) between 
the PEs and instead requires explicit buffered transfer or copying of the product terms (for multiply-accumulate) between BRAM columns.  
PiCaSO enables zero-copy reduction operations with the operand-multiplexer (OpMux) shown in Fig.~\ref{fig:archPIM}.
The operand-multiplexer allows pass-through of bitlines from BRAMs to ALUs for multiplication
but then supports zero-copy reduction summation of the product terms.
The Network Node in Fig.~\ref{fig:archPIM} provides a streaming interface between PIM blocks 
enabling the streaming of partial products into the ALU of the destination PE for summation, without intermediate copying.  
Section~\ref{sec:opmux} presents how the reduction operation can be optimized by inserting 
pipeline stages that overlap data transfers with ALU operations, hiding the transfer latency. 

\begin{table}
\renewcommand{\arraystretch}{0.9}
\caption{Full Adder/Subtractor (FA/S) Op-Codes}
\label{tab:aluOp}
\centering
\begin{tabular}{ccl}
\hline
Op-Code & Output (SUM) & \multicolumn{1}{c}{Description} \bigstrut[t]\\
\hline
ADD    & X + Y    & Acts as a Full-Adder (FA) \bigstrut[t]\\
SUB    & X - Y    & Acts as an FA with borrow logic \\
CPX    & X        & Copies operand X unmodified \\
CPY    & Y        & Copies operand Y unmodified \\
\hline
\end{tabular}%
\end{table}

\begin{table}
\renewcommand{\arraystretch}{0.9}
\caption{Op-Encoder Configurations for Booth's Radix-2 Multiplier}
\label{tab:opEncoder}
\centering
\begin{tabular}{cccc}
\hline
Conf   & YX     & ALU Op-Code & Description \bigstrut[t]\\
\hline
000    & xx     & ADD    & Request ADD \bigstrut[t]\\
001    & xx     & CPX    & Select X operand \\
010    & xx     & CPY    & Select Y operand \\
011    & xx     & SUB    & Request SUB \\
\hline
1xx    & 00     & CPX    & NOP \bigstrut[t]\\
1xx    & 01     & ADD    & +Y \\
1xx    & 10     & SUB    & -Y \\
1xx    & 11     & CPX    & NOP \\
\hline
\end{tabular}%
\end{table}

\subsection{Parallel to Serial Corner Turning}

PiCaSO is a bit-serial array processor designed to work with standard processors.  
Parallel data read/written from DRAM and external I/O devices is corner-turned into bit-serial data and stored 
as a striped column within the BRAMs.  
This is a standard storage scheme for bit serial ALUs similar to~\cite{eriko,comefa,spar22}.
PiCaSO configures a BRAM to be 16-bits wide to concurrently feed bit-serial data to 16 ALUs~\cite{spar22}. 
In SPAR-2~\cite{spar22}, the benchmark overlay, the 16 PEs form a logical 4$\times$4 PE Block. 
PiCaSO structurally organizes the PE-Block as a 1$\times$16 linear array to optimize 
layout in the columnar architecture of Virtex FPGAs.  
This reduces routing complexity and wire delay, allowing a greater number of PEs to be synthesized into 
the FPGA and improving system clock speed. 

\subsection{Bit-Serial ALUs} 
\label{sec:alu}

Fig.~\ref{fig:archPIM}(b) shows the architecture of the bit-serial ALU consisting of a Full-ADD/SUB module (FA/S) and an op-code encoder.
The FA/S implements the four operations in Table~\ref{tab:aluOp}.
CPX and CPY support min/max pooling and other
filter operations that require the selection of one of the two input operands.
Op-Encoder provides an abstract interface for the FA/S module. 
Table~\ref{tab:opEncoder} shows the encoding for Booth's Radix-2 multiplication algorithm. 

\subsection{Supporting Reduction Operations}
\label{sec:opmux}
\begin{figure}
\centering
\includegraphics[width=\linewidth]{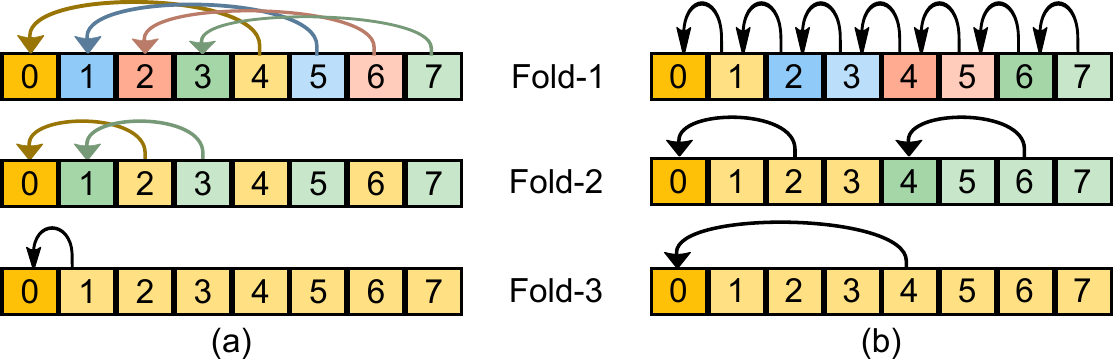}
\caption{Folding patterns in Operand Multiplexer (OpMux)}
\label{fig:folding}
\end{figure}

The operand-multiplexer (OpMux) provides a data path for reduction operations between the PEs in a PE-Block without having to copy the operands between bitlines. 
This is achieved using a folding technique.
Fig.~\ref{fig:folding} shows two types of folding patterns for a PE row with 8 columns enabled by the OpMux module.
In pattern (a), after adding an operand with its 
fold-1 pattern, PE 0, 1, 2, and 3 contain the summation of
0 \& 4, 1 \& 5, 2 \& 6, and 3 \& 7 respectively.
In pattern (b), after adding an operand with its fold-1 pattern,
PE 0, 2, 4, and 6 contain the summation of 0 \& 1, 2 \& 3, 4 \& 5, and 6 \& 7.
In both cases, after applying fold-1, fold-2, and fold-3 in that order,
the accumulation result will be stored in PE-0.
Fold-1 of the pattern (b) can be especially useful in CNN models, where
each PE needs access to its adjacent PEs. 
A similar type of folding scheme can be realized using multiplexers at the output of SAs in custom PIM blocks. 
Results presented in Section~\ref{sec:archfeature} show the potential reduction in accumulation 
latency for the custom designs provided with this optimization.

\begin{table}
\begin{threeparttable}[b]  
\renewcommand{\arraystretch}{0.8}
\setlength{\tabcolsep}{4pt}
\caption{Configurations of Operand Multiplexer}
\label{tab:opmuxConf}
\centering
\begin{tabular}{cccl}
\hline
Config Code & X      & Y      & \multicolumn{1}{c}{Description} \bigstrut [t]\\
\hline
A-OP-B & A        & B            & Used in standard operations \bigstrut[t]\\
A-FOLD-1 & A      & \{0, A[H2]\} & A[H2]: second half of A \\
A-FOLD-2 & A      & \{0, A[Q2]\} & A[Q2]: second quarter of A \\
A-FOLD-3 & A      & \{0, A[HQ2]\} & A[HQ2]: second half-quarter of A \\
A-FOLD-4 & A      & \{0, A[HHQ2]\} & A[HHQ2]: second half of A[HQ1]\tnote{1} \\
A-OP-NET & A      &  NET         & Operates on network stream \\
0-OP-B & 0        & B      & Used in the first iteration of MULT \\
\hline
\end{tabular}%
\begin{tablenotes}
\item[1] A[HQ1] : first half-quarter of A \\
\end{tablenotes}
\end{threeparttable}
\vspace{-5pt}
\end{table}

Table~\ref{tab:opmuxConf} shows configurations currently supported by the OpMux module.
Configuration A-OP-B connects ports A to X and B to Y and is used in element-wise operations.
A-FOLD-x implements folding patterns similar to Fig.~\ref{fig:folding}(a).
A-OP-NET directly feeds the network stream into the ALU.
0-OP-B is used as the initialization step in Booth's multiplication.

\begin{figure}[t]
\centering
\includegraphics[width=\linewidth]{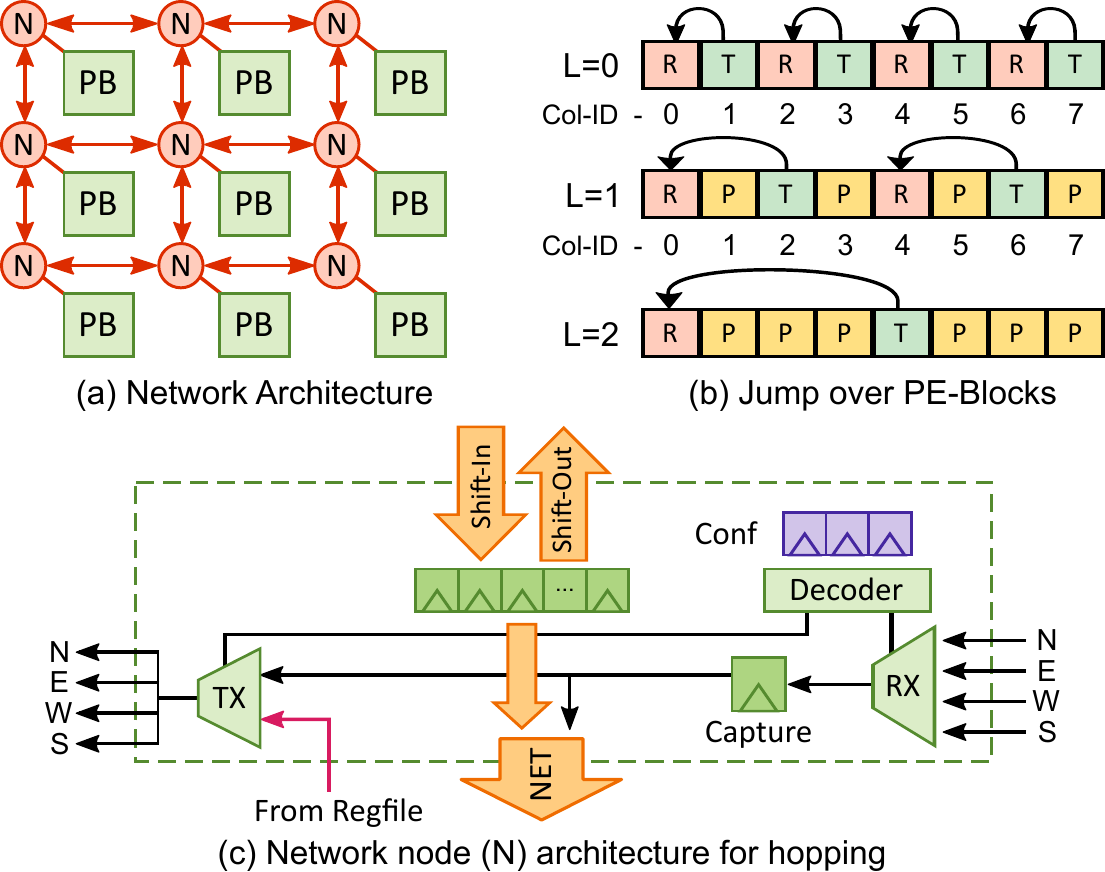}
\caption{Data network for fast accumulation and reduction operations}
\label{fig:archNet}
\end{figure}

\begin{table*}
\renewcommand{\arraystretch}{0.8}
\setlength{\tabcolsep}{3pt}
\caption{Comparison between tiles of $4\times4$ PE-Blocks of different overlay configurations}
\label{tab:tile4x4}
\centering

\begin{tabular}{|c|cc|cc|cc|cc|cc|cc|cc|cc|cc|cc|}
\hline
       & \multicolumn{4}{c|}{\Agraycell Benchmark~\cite{spar22}} & \multicolumn{4}{c|}{Full-Pipe}    & \multicolumn{4}{c|}{\Agraycell Single-Cycle} & \multicolumn{4}{c|}{RF-Pipe}      & \multicolumn{4}{c|}{\Agraycell Op-Pipe} \bigstrut[t]\\
\hhline{~|*{20}{-}}  & \multicolumn{2}{c|}{\Agraycell Virtex-7} & \multicolumn{2}{c|}{\Agraycell U55} & \multicolumn{2}{c|}{Virtex-7} & \multicolumn{2}{c|}{U55} & \multicolumn{2}{c|}{\Agraycell Virtex-7} & \multicolumn{2}{c|}{\Agraycell U55} & \multicolumn{2}{c|}{Virtex-7} & \multicolumn{2}{c|}{U55} & \multicolumn{2}{c|}{\Agraycell Virtex-7} & \multicolumn{2}{c|}{\Agraycell U55} \bigstrut[t]\\
       & \Agraycell Tile & \Agraycell Block & \Agraycell Tile & \Agraycell Block & Tile   & Block  & Tile   & Block  & \Agraycell Tile & \Agraycell Block & \Agraycell Tile & \Agraycell Block & Tile   & Block  & Tile   & Block  & \Agraycell Tile & \Agraycell Block & \Agraycell Tile & \Agraycell Block \\
\hline
LUT    & \Agraycell 3023 & \Agraycell 189 & \Agraycell 2449 & \Agraycell 153 & 835    & 52     & 774    & 48     & \Agraycell 895 & \Agraycell 56 & \Agraycell 1068 & \Agraycell 67 & 1017   & 64     & 1064   & 67     & \Agraycell 836 & \Agraycell 52 & \Agraycell 774 & \Agraycell 48 \bigstrut[t]\\
FF     & \Agraycell 1024 & \Agraycell 64 & \Agraycell 768 & \Agraycell 48 & 1799   & 112    & 1799   & 112    & \Agraycell 1031 & \Agraycell 64 & \Agraycell 1031 & \Agraycell 64 & 1543   & 96     & 1527   & 95     & \Agraycell 1543 & \Agraycell 96 & \Agraycell 1543 & \Agraycell 96 \\
Slice  & \Agraycell 1056 & \Agraycell 66 & \Agraycell 556 & \Agraycell 35 & 522    & 33     & 243    & 15     & \Agraycell 395 & \Agraycell 25 & \Agraycell 223 & \Agraycell 14 & 451    & 28     & 243    & 15     & \Agraycell 472 & \Agraycell 30 & \Agraycell 295 & \Agraycell 18 \\
\hline
Max-Freq & \multicolumn{2}{c|}{\Agraycell 240 MHz} & \multicolumn{2}{c|}{\Agraycell 445 MHz} & \multicolumn{2}{c|}{540 MHz} & \multicolumn{2}{c|}{737 MHz} & \multicolumn{2}{c|}{\Agraycell 245 MHz} & \multicolumn{2}{c|}{\Agraycell 487 MHz} & \multicolumn{2}{c|}{360 MHz} & \multicolumn{2}{c|}{600 MHz} & \multicolumn{2}{c|}{\Agraycell 370 MHz} & \multicolumn{2}{c|}{\Agraycell 620 MHz} \bigstrut[t]\\
\hline
\end{tabular}%

\vspace{-1.5pt}
\end{table*}

\subsection{Network Architecture} \label{sec:netArch}
Fig.~\ref{fig:archNet}(c) expands the Network Node shown in Fig.~\ref{fig:archPIM}. 
Fig.~\ref{fig:archNet}(a) shows the PE-Blocks (PB) connected to the data network through the network module (N).
Fig.~\ref{fig:archNet}(b) illustrates data reduction patterns between 8 nodes.
Each node can be configured as a 
transmitter (T), receiver (R), or pass-through (P) based on a level (L) parameter and its position in the array.
Fig.~\ref{fig:archNet}(b), shows that level 0 logically connects even nodes as receivers with 
their right neighbor as transmitters between columns.
For level 1, the middle node of every 3 consecutive nodes acts as a pass-through,
effectively connecting its neighbors. 
Similarly, level~2 connects node-4 to node-0. 
During accumulation, bits of the operand in the transmitter hop through P-nodes to reach the receiver ALU
where they are added (serially) to the operand in the receiver. 
After levels 0, 1, and 2, PE 0 contains the accumulation result of an entire row in the array.

\subsection{Pipelining Options for PIM-Blocks}
\label{sec:d_pipeline}
The dashed registers in Fig.~\ref{fig:archPIM}(a) show three potential points for pipelining the PIM Block: 
register file output, OpMux output, and ALU output.
The \emph{Single-Cycle} configuration has no pipeline stages and is 
equivalent to the custom BRAM designs~\cite{comefa,eriko} and the benchmark overlay~\cite{spar22}.
PiCaSO can be configured in different pipeline configurations based on network requirements and choice of FPGA.  
\emph{RF-Pipe} inserts a pipeline stage at the register file output to hide the read latencies of the BRAM.
\emph{Op-Pipe} inserts a pipeline stage at the OpMux output to hide long wire delays through the network.
\emph{Full-Pipe}, referred to as PiCaSO-F, enables all three pipeline stages as shown in Fig.~\ref{fig:archPIM}~(a).   

\section{Analysis}
\label{sec:analysis}

\subsection{Performance and Utilization} \label{sec:result-block}
Table~\ref{tab:tile4x4} compares the pipeline configurations outlined in subsection~\ref{sec:d_pipeline} against 
SPAR-2, the benchmark overlay from~\cite{spar22}.  
All designs were implemented and run on Virtex-7 (xc7vx485) and Alveo U55 FPGAs.
Utilization numbers follow the tile definition in SPAR-2 consisting of 256 PEs organized 
in a 4$\times$4 array of PE blocks, with 16 PEs in each block.
The total utilization per tile and the average utilization per block are shown.
The Full-Pipe configuration achieved a 2.25$\times$ and a 1.67$\times$
increase in clock frequency compared to the benchmark design on Virtex-7 and U55 devices, respectively.
In both devices, Full-Pipe provided a 2$\times$ improvement in resource utilization over SPAR-2.

\begin{table}
\begin{threeparttable}
\renewcommand{\arraystretch}{0.8}
\setlength{\tabcolsep}{4pt}
\caption{Cycle latency of different operations}
\label{tab:cycles}
\centering
\begin{tabular}{ccc}
\hline
Operation & Benchmark~\cite{spar22} & PiCaSO-F \bigstrut[t]\\
\hline
ADD/SUB               &    $2N$           &   $2N$ \bigstrut[t]\\
MULT\tnote{1}         &    $2N^2 + 2N$    &   $2N^2 + 2N$ \\
Accumulation\tnote{2}    &   $(q - 1 + 2 \log_2 q) N$    & $15 + \frac{q}{16} + 4N + (N+4)J$ \\
$q$ = 128, $N$ = 32    &   4512            & 259 \\
\hline
\end{tabular}%
\begin{tablenotes}
\item[1] Booth's Radix-2 multiplication 
\item[2]
\Abox{$q$} : Number of columns to be accumulated \\
\Abox{$N$} : Operand width \\
\Abox{$J$} : Number of network jumps needed = $\log_2 (q/16)$\\
\end{tablenotes}
\end{threeparttable}
\vspace{-8pt}
\end{table}

The Single-Cycle configuration achieved similar performance on the Virtex-7 and better performance on the U55 compared to the benchmark system, 
with 2.6$\times$ and 2.5$\times$ utilization improvements, respectively.
It had a smaller flip-flop count and slice utilization compared to the Full-Pipe due to the absence of the pipeline registers.
Both RF-Pipe and Op-Pipe achieved better clock speeds 
but with an increase in slice utilization compared to Single-Cycle, due to the addition of the pipeline stages.
As argued in Subsection~\ref{sec:d_pipeline}, Op-Pipe had 
better performance compared with RF-Pipe by minimizing the clock latency contributed by the network.
All configurations offered at least 2$\times$ better utilization 
and up to 2$\times$ better performance compared to the benchmark design.

Table~\ref{tab:tile4x4} shows Full-Pipe achieved clock frequencies of 540~MHz on the Virtex-7 (xc7vx485-2), 
and 737~MHz on the Alveo U55 (xcu55c, -2 speed grade).
The data sheets for these devices list 543.77~MHz and 737~MHz, respectively as the maximum BRAM clock frequencies.
Surprisingly, this is an improvement over the custom designs reported in~\cite{eriko, comefa}. 
The technology node of U55 (16~nm) is comparable to that of the designs proposed in CCB (Stratix 10, 14~nm) and CoMeFa (Arria 10, 20~nm).
Yet, PiCaSO runs 1.62$\times$ and 1.25$\times$ faster than the fastest configurations of CCB and CoMeFa, respectively.
This is due to the pipelined architecture of PiCaSO, where the slowest stage is the BRAM.
Thus, it can run as fast as the maximum frequency of the BRAM.

\subsection{Reduction Network}

Both PiCaSO and SPAR-2~\cite{spar22} use Booth's Radix-2 algorithm for multiplication. 
Thus, the cycle latencies for the ADD/SUB and MULT operations in Table~\ref{tab:cycles} are identical. 
SPAR-2 uses a standard NEWS network to copy operands between PEs when summing the partial products during multiply-accumulate (MAC) operations.
The Accumulation row compares the number of clock cycles for SPAR-2's NEWS network and PiCaSO's reduction network.
The last row in Table~\ref{tab:cycles} shows the PiCaSO-F reduction network provides 
a 17$\times$ improvement in accumulation latency for the test configuration reported in ~\cite{spar21}.
This improvement is due to the careful design of the binary-hopping network discussed in Section~\ref{sec:netArch}, 
which overlaps data transfer with computation during accumulation.

\begin{table}
\renewcommand{\arraystretch}{0.9}
\setlength{\tabcolsep}{3pt}
\caption{Comparison of largest overlay arrays in Virtex devices}
\label{tab:largest}
\centering
\begin{tabular}{c|cc|cc}
\hline
       & \multicolumn{2}{c|}{Virtex-7} & \multicolumn{2}{c}{Alveo U55} \bigstrut[t]\\
       & Benchmark~\cite{spar22} & PiCaSO-F & Benchmark~\cite{spar22} & PiCaSO-F \\
\hline
Max-Size & 24K    & 33K    & 63K    & 64K \bigstrut[t]\\
LUT    & 74.6\% & 32.5\% & 41.6\% & 14.8\% \\
FF     & 16.0\% & 38.0\% & 9.7\%  & 17.3\% \\
BRAM   & 73.8\% & 99.9\% & 98.4\% & 100.0\% \\
Uniq. Ctrl. Set & 32.1\% & 2.1\%  & 19.5\% & 0.8\% \\
Slice  & 86.0\% & 76.4\% & 63.4\% & 32.0\% \\
\hline
\end{tabular}%
\end{table}

\subsection{Scalability}

A primary design goal for PiCaSO was to make it scale linearly with the BRAM 
capacity of any FPGA.
To evaluate scalability, the largest-sized array of PIM blocks that could fit into the target devices was constructed.
The results of this study are shown in Table~\ref{tab:largest}.

In the Virtex-7 FPGA, the largest array of SPAR-2~\cite{spar22} PIM blocks contained 24K PEs.
This did not achieve the full capacity of the Slices or BRAM resources available in that device.
The implementation tool failed at the placement step for larger arrays due to a high utilization (32.1\%) of Unique Control Sets.
Control sets are the collection of control signals for slice flip-flops.
Flip-flops must belong to the same control set to be packed into the same slice.
A large number of unique control sets makes it difficult to find a valid placement, even with enough available slices.
In contrast, PiCaSO-F fully utilized the BRAM resources to fit 33K PEs, a 37.5\% improvement over SPAR-2 in the same device.
PiCaSO does not suffer from the placement issues observed in SPAR-2 due to a very low (2.1\%) utilization of the unique control sets.

In the U55 FPGA, SPAR-2 almost achieved the full BRAM capacity for an array size of 63K PEs.
This is due to the U55 FPGA offering significantly more slices and routing resources compared to the Virtex-7 FPGA.
PiCaSO achieved 100\% utilization of BRAM with 2$\times$ better slice utilization over SPAR-2.

Our results showed that the scalability of the benchmark design, SPAR-2, is dependent on the Slice-to-BRAM ratio and 
cannot guarantee the creation of a PIM array that scales with the BRAM capacity.  
Conversely, our results showed PiCaSO scaling with the BRAM capacity independent of the Slice-to-BRAM 
ratio across multiple devices of Virtex-7 and Ultrascale+ FPGA families.
Table~\ref{tab:devices} lists representative devices 
we evaluated based on the following two criteria: BRAM capacity and LUT-to-BRAM ratio.
Each device is assigned an ID as a short name to be used in this paper.

\begin{figure}
\centering
\includegraphics[width=\linewidth]{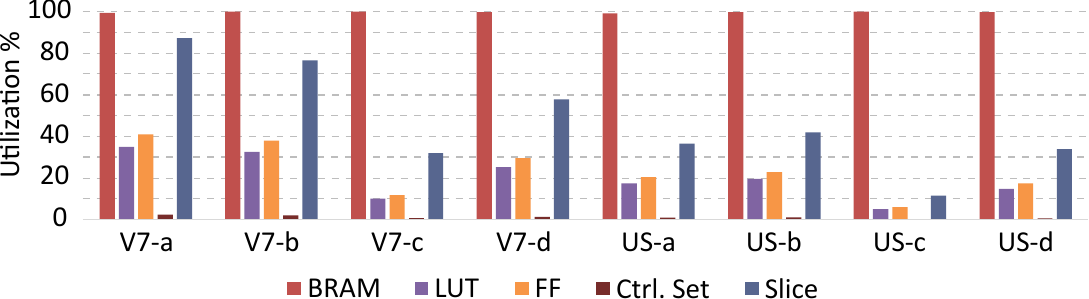}
\caption{Scalability study on Virtex-7 and Ultrascale+ FPGA families}
\label{fig:scalability}
\end{figure}

Fig.~\ref{fig:scalability} shows that PiCaSO utilized the full BRAM capacity in all devices,
and achieved the maximum number of PEs the device can fit based on BRAM density. 
Results showed for the smallest device (V7-a) and lowest LUT-to-BRAM ratio, the LUT and flip-flop utilization is around 40\%.  
For one of the largest devices with a high LUT-to-BRAM ratio (US-c), these utilization numbers are negligible, around 5\%.
These results strongly support that PiCaSO scales linearly with the BRAM capacity of the device.

\begin{table}
\begin{threeparttable}
\caption{Representative of Virtex-7 and Ultrascale+ devices}
\label{tab:devices}
\centering
\begin{tabular}{cccccc}
\hline
Device & Tech   & BRAM\# & Ratio\tnote{1} & Max PE\#\tnote{2}  & ID \bigstrut[t] \\
\hline
xc7vx330tffg-2 & V7     & 750    & 272    & 24K      & V7-a \bigstrut[t]\\
xc7vx485tffg-2 & V7     & 1030   & 295    & 32K      & V7-b \\
xc7v2000tfhg-2 & V7     & 1292   & 946    & 41K      & V7-c \\
xc7vx1140tflg-2 & V7    & 1880   & 379    & 60K      & V7-d \\
xcvu3p-ffvc-3 & US+     & 720    & 547    & 23K      & US-a \\
xcvu23p-vsva-3 & US+    & 2112   & 488    & 67K      & US-b \\
xcvu19p-fsvb-2 & US+    & 2160   & 1892   & 69K      & US-c \\
xcvu29p-figd-3 & US+    & 2688   & 643    & 86K      & US-d \\
\hline
\end{tabular}%
\begin{tablenotes}
\item[1] LUT-to-BRAM ratio
\item[2] Maximum number of PEs if all BRAMs are utilized
\end{tablenotes}
\end{threeparttable}
\end{table}

\begin{figure}[t]
\centering
\includegraphics[width=\linewidth]{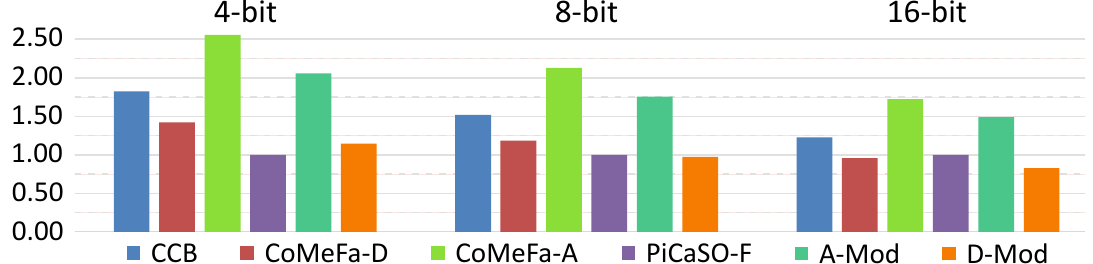}
\caption{Relative MAC latency of custom designs w.r.t PiCaSO}
\label{fig:maclatency}
\end{figure}

\begin{figure}
\centering
\includegraphics[width=\linewidth]{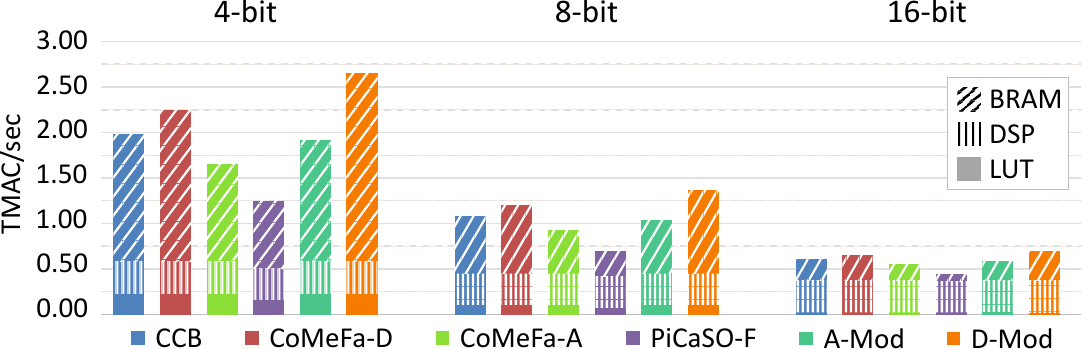}
\caption{Peak MAC throughput of PiCaSO and custom designs on Alveo U55}
\label{fig:throughput}
\end{figure}

\begin{figure}[!ht]
\centering
\includegraphics[width=\linewidth]{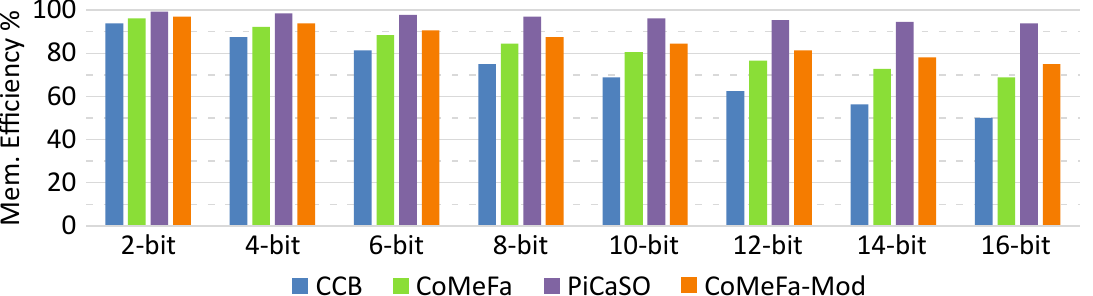}
\caption{BRAM memory utilization efficiency on Virtex devices}
\label{fig:memutil}
\end{figure}

\section{Comparison with Custom Designs}
\label{sec:performance}
Fig.~\ref{fig:maclatency} shows the relative MAC latency of the custom designs w.r.t PiCaSO.
The latency is computed for 16 parallel MULTs followed by the accumulation of the products.
The clock speeds of custom designs are adjusted based on the performance degradations reported in~\cite{eriko, comefa}.
With the exception of CoMeFa-D at 16-bit precision, PiCaSO has the shortest latency due to faster clock speed and accumulation.
CCB and CoMeFa extend the clock period to allow a complete read-modify-write per clock cycle. 
This allows a complete MULT to finish in half the number of cycles compared to PiCaSO and can reduce latencies at higher precisions.
Still, PiCaSO runs 1.72$\times$ -- 2.56$\times$ faster than CoMeFa-A, which is reported as the most practical design in~\cite{comefa}.

Peak TeraMAC/sec throughputs on the U55 FPGA are shown in Fig.~\ref{fig:throughput}.
CCB and CoMeFa design the BRAM IP to support one PE per bitline.
With a column muxing factor of 4~\cite{comefa}, a Virtex 36Kb BRAM would be redesigned as a 256$\times$144 array with 144 PEs per BRAM.
The use of standard BRAM IP prevents PiCaSO (and all overlays) from making this modification. 
Yet PiCaSO still 
achieves 75\% -- 80\% of CoMeFa-A's peak throughput, the most practical of the two CoMeFa designs. 
This results from PiCaSO not sacrificing the same degradation of clock speed seen in all custom designs.

The memory utilization efficiency of BRAMs is not discussed in~\cite{eriko,comefa} but we feel is an important metric for PIM architectures.
Memory utilization efficiency can be defined as the fraction of BRAM memory that can be used to store model weights.
Both CCB and CoMeFa follow the computation techniques used in~\cite{eckert}
which requires scratchpad memory.
For N-bit operands, CCB requires 8N reserved wordlines.
CoMeFa only needs 5N wordlines using the ``One Operand Outside RAM (OOOR)" technique.
PiCaSO requires only 4N wordlines, as it does not require copying operands to the same bitline as in CoMeFa. 
In the widest mode of a Virtex 36Kb BRAM, each PE of CCB and CoMeFa would have 256 bits of storage in 
its register file (bitline). 
For PiCaSO, each register file has 1024 bits.
Fig.~\ref{fig:memutil} shows the memory utilization efficiency of these architectures.
As observed, 
at higher precisions the memory efficiency drops significantly for CCB and CoMeFa.
For 16-bit operands, CCB and CoMeFa have only 50\% and 68.8\% efficiencies, respectively,
while PiCaSO has 93.8\% efficiency.

\label{sec:archfeature}

Table~\ref{tab:customPIM} summarizes comparisons between PiCaSO and the custom designs.
The custom designs significantly degrade the BRAMs maximum clock frequency, whereas PiCaSO runs at the maximum clock speed of the BRAM.
However, PiCaSO has 1/4\textsuperscript{th} the number of parallel MACs, as it cannot access all the bitlines. 
Multiplication in PiCaSO is 2$\times$ slower, as it requires 2 cycles to process a single bit.  
However, accumulation is 2$\times$ faster in PiCaSO.
PiCaSO supports Booth's radix-2 multiplication.
In Booth's algorithm, half of the intermediate steps are NOPs on average.
Thus, PiCaSO can potentially further reduce the multiplication latency by 50\% on average. 
CoMeFa can use Booth's algorithm only in OOOR mode and CCB does not support it at all.

As discussed earlier, the memory utilization efficiency of CCB is 
significantly low, PiCaSO is high, and CoMeFa lies in between.
CCB has the highest design complexity mainly due to its need for a modified voltage supply.
CoMeFa has medium complexity since it requires modifications to the SAs,
additional flip-flops, and SA cycling.
Being an overlay, PiCaSO does not have such design complexities.
As reported in~\cite{comefa}, the practicality of CCB is low, CoMeFa-D is medium, and CoMeFa-A is high.
In that reference, the practicality of PiCaSO is very high.
It offers 80\% of CoMeFa-A's peak throughput with 2.56$\times$ shorter latency, 25\% better memory efficiency, 
can be implemented using off-the-shelf FPGAs, and is tested on real devices,
while CCB and CoMeFa numbers are mainly based on simulations.

\begin{table}
\begin{threeparttable}
\centering
\caption{Comparison with Customized BRAM PIM architectures}
\label{tab:customPIM}
\setlength{\tabcolsep}{3pt}
\begin{tabular}{lccccc}
\hline
       & CCB    & CoMeFa-D & CoMeFa-A & PiCaSO-F & A-Mod \bigstrut[t]\\
\hline
Architecture & Custom & Custom & Custom & Overlay & Custom \bigstrut[t]\\
Clock Overhead & 60\%   & 25\%   & 150\%  & 0\%    & 150\% \\
Parallel MACs & 144    & 144    & 144    & 36     & 144 \\
\hline
Mult Latency\tnote{1} & (a)    & (a)    & (a)    & (b)    & (a) \bigstrut[t]\\
N = 8  & 86     & 86     & 86     & 144    & 86 \\
\hline
Accum. Latency\tnote{2} & (c)    & (c)    & (c)    & (d)    & (e) \bigstrut[t]\\
q = 16, N = 8 & 80     & 80     & 80     & 48     & 40 \\
\hline
Support Booth's & No     & Partial & Partial & Yes    & Yes \bigstrut[t]\\
Mem. Efficiency & Low    & Medium & Medium & High   & Medium \\
Complexity & High   & Medium    & Medium & No     & Medium \\
Practicality & Low    & Medium & High   & Very High & High \\
\hline
\end{tabular}%
\begin{tablenotes}
\item[1] (a) $N^2 + 3N -2$ ; \ (b) $2N^2 + 2N$
\item[2] (c) $(2N + \log_2 q) \log_2 q$ ; \ (d) $(N+4) \log_2 q$ ; \ (e) $(N+2) \log_2 q$
\end{tablenotes}
\end{threeparttable}
\end{table}

\subsection{Fusing PiCaSO Optimizations into Custom Designs }

\begin{figure}
\centering
\includegraphics[width=\linewidth]{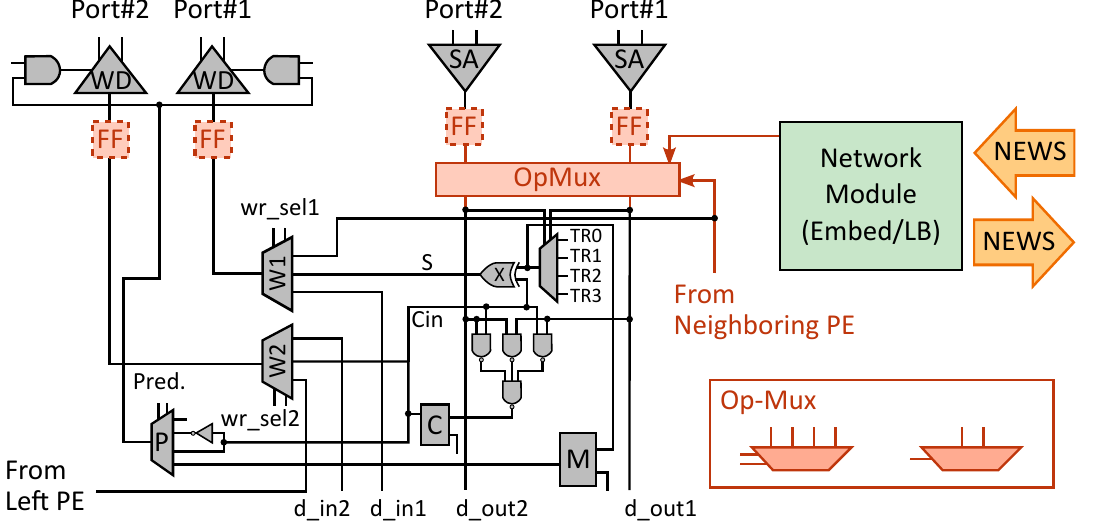}
\caption{Modified CoMeFa-A~\cite{comefa} with PiCaSO adoption (A-Mod)}
\label{fig:ComefaMod}
\end{figure}

Fig.~\ref{fig:ComefaMod} shows how modifications highlighted in red can accelerate CoMeFa-A~\cite{comefa}.
We refer to this implementation as \emph{A-Mod}.
PiCaSO's OpMux module per bitline consists of a 2-to-1 mux and a 4-to-1 mux.
This can be implemented using a few CMOS pass transistors. 
OpMux then saves both the cycles and memory needed to copy operands during accumulation~\cite{eckert,eriko}. 
PiCaSO's network module can overlap data movement with computation between different PIM blocks. 
The network module can be embedded within the PIM block or can be implemented using logic slices from the FPGA.
A single-bit port connection to the network module is enough to support row-wise accumulation.

Although~\cite{comefa} mentions that the PE does not add additional delay to the extended clock,
in a practical circuit, there will always be an additional delay.
This delay can be hidden using one of the pipelining schemes of PiCaSO.
A single stage of registers could be enough to hide the PE delay.
As BRAM blocks already contain output registers, this should not add any additional
area overhead on top of what is reported in~\cite{comefa}.
The PE circuit can be placed between two stages of registers if the delay is too long.
This is illustrated using the dashed flip-flops in Fig.~\ref{fig:ComefaMod}.
Similar modifications can be performed on CoMeFa-D referred to as implementation \emph{D-Mod}.

These modifications can significantly improve the performance of the custom designs.
The extrapolated performance numbers for A-Mod and D-Mod are presented in Fig.~\ref{fig:maclatency}~and Fig.~\ref{fig:throughput}.
As observed in Fig.~\ref{fig:maclatency}, the adoption of PiCaSO's OpMux and network modules can improve 
their MAC latency by \mbox{13.4\% -- 19.5\%} due to faster accumulation.
This consequently improves their throughput by 5\% - 18\% over different precisions.
In Fig.~\ref{fig:memutil}, CoMeFa-Mod represents both A-Mod and D-Mod implementations.
Due to OpMux, A-Mod and D-Mod no longer requires scratchpad storage to copy operands for accumulation.
This improves their memory utilization efficiency by 6.2\%.
This means at 4-bit precision, 1.6 million more weights can be stored in a device with 100~Mb of BRAM. 
This would significantly reduce weight stall cycles~\cite{tpu2017} and allow bigger models to be stored on chip.
In Table~\ref{tab:customPIM}, the A-Mod column shows the architectural enhancements due to these modifications.
A-Mod retains the high parallelism and fast Mult latency of the original CoMeFa design and 
offers 2$\times$ faster accumulation and full support for Booth's algorithm.

\section{Conclusions}
\label{sec:conclusion}
This paper presented PiCaSO, an open-source scalable and portable Processor in Memory (PIM) overlay architecture.  
As an overlay, PiCaSO brings software levels of productivity to the design of FPGA machine-learning accelerators across AMD devices.   
The PIM architecture addresses the needs of machine learning and big data analytic applications that are memory intensive.  
  
A scalability study was presented that established PiCaSO scaled linearly with the BRAM capacity across a range of devices with varying LUT-to-BRAM ratios.  
Analysis on \mbox{Virtex-7} and Ultrascale+ devices showed PiCaSO runs as fast as the BRAM maximum frequency.  
Comparisons against SPAR-2, a state-of-the-art SIMD array processor overlay, showed improvements in slice utilization and achievable clock frequency by 2$\times$ and accumulation latency reduction by 17$\times$.

Comparative analysis against custom designs showed PiCaSO achieves up to 80\%
of the peak throughput and up to 2.56$\times$ shorter latency and 25\% -- 43\%
better memory utilization.  

We showed that the proposed architecture can be adopted into custom PIM designs,
and can improve the throughput by 18\%, latency by 19.5\%, and memory utilization by 6.2\%.

Our future efforts are focused on automating and applying application-specific 
and logic-family customizations to the generation of both PiCaSO-based accelerator and compiler-generated executables.

\bibliographystyle{IEEEtran}
\IEEEtriggeratref{5}
\IEEEtriggercmd{\balance}
\bibliography{IEEEabrv,ref}

\end{document}